\documentclass[aps,prd,preprint,superscriptaddress,amsmath,amssymb,showpacs]{revtex4-1}
\usepackage{dcolumn}
\usepackage{graphicx}
\usepackage{float}
\usepackage{physics}
\usepackage[colorlinks=true,allcolors=blue]{hyperref}

\begin{document}
	
	\title{Shear viscosity coefficient of magnetized QCD medium with anomalous magnetic moments near chiral phase transition}
	
	\author{Yi-Wei Qiu}
	\affiliation{College of Science, China Three Gorges University, Yichang 443002, China}
	
	\author{Sheng-Qin Feng}
	\email{Corresponding author: fengsq@ctgu.edu.cn}
	\affiliation{College of Science, China Three Gorges University, Yichang 443002, China}
	\affiliation{Center for Astronomy and Space Sciences and Institute of Modern Physics, China Three Gorges University, Yichang 443002, China}
	\affiliation{Key Laboratory of Quark and Lepton Physics (MOE) and Institute of Particle Physics,\\
		Central China Normal University, Wuhan 430079, China}
	
	\author{Xue-Qiang Zhu}
	\affiliation{Key Laboratory of Quark and Lepton Physics (MOE) and Institute of Particle Physics,\\
		Central China Normal University, Wuhan 430079, China}
	\date{\today}
	
	\begin{abstract}
		Abstract:  We study the properties of the shear viscosity coefficient of quark matter near the chiral phase transition at finite temperature and chemical potential, and the kinds of high temperature, high density and strong magnetic field background.  The strong magnetic field induces anisotropy, that is, the quantization of Landau energy levels in phase space. If the magnetic field is strong enough, it will interfere with significant QCD phenomena, such as the generation of dynamic quark mass, which may affect the transport properties of quark matter. The inclusion of the anomalous magnetic moments of the quarks at finite density into the Nambu-Jona-Lasinio model gives rise to additional spin polarization magnetic effects. It is found that both the ratio $\eta/s$ of shear viscosity coefficient to entropy and the collision relaxation time $\tau$ show  similar trend with temperature, both of which reach minima around the critical temperature. The shear viscosity coefficient of the dissipative fluid system can be decomposed into five different components as the strong magnetic field exists. The influences of the order of chiral phase transition and the critical end point on dissipative phenomena in such a magnetized medium are quantitatively investigated. It is found that ${\eta}_{1}$, ${\eta}_{2}$, ${\eta}_{3}$, and ${\eta}_{4}$ all increase with temperature. For first-order phase transitions, ${\eta}_{1}$, ${\eta}_{2}$, ${\eta}_{3}$, and ${\eta}_{4}$ exhibit discontinuous characteristics.
	\end{abstract}
	
	
	\maketitle
	
	\section{Introduction}\label{sec:01_intro}
	
	The study of the strong interaction under the influence of different external agents, including strong magnetic field \cite{Bzdak:2011yy,Skokov:2009qp,Deng:2012pc,Kharzeev:2007jp,Zhong:2014cda,Mo:2013qya,Feng:2016srp} and large angular momentum rotation \cite{Liang:2004ph,Gao:2007bc,Guo:2019joy,Becattini:2007sr,Zhao:2022uxc,Zhao:2023pne} will eventually shed light on the complexity of its phase diagram and transport properties, or could reveal new unknown features. Primarily due to the complexity of fundamental theories, the study of low energy states relies on specific tools such as lattice simulations and effective models. The Nambu-Jona-Lasinio (NJL) model has been proven to be a useful conceptual tool for solving different problems \cite{Miransky:2015ava,Andersen:2014xxa}. As a particular chapter of this method, one can mention the study of quarks interacting with external magnetic fields, where a variety of issues have been studied, such as magnetic catalysis \cite{Gusynin:1994xp}, quark matter transport characteristics \cite{Zhuang:1995uf,Zhu:2022irg,Ghosh:2018cxb}, magnetic oscillations \cite{Ebert:1999ht}, vector \cite{Denke:2013gha} and tensor \cite{Ferrer:2013noa,Qiu:2023kwv} additional couplings.
	
	In a magnetic field environment, it is naturally believed that the chiral critical temperature will increase with the magnetic field, which is known as the magnetic catalytic (MC) effect. However, lattice QCD results \cite{Bali:2011qj,Bornyakov:2013eya} detected the feature of inverse magnetic catalytic effect (IMC) near the critical temperature. Recent studies \cite{Qiu:2023kwv,Xu:2020yag,Chaudhuri:2019lbw} have shown that the addition of the anomalous  magnetic  moments (AMM) can also exhibit IMC properties consistent with the lattice QCD results. The remaining IMC puzzle and recently discovered properties of magnetized matter attract our renewed interest to revisit QCD vacuum and matter under an external magnetic field and to find the underlying mechanism for these properties. It is known that the dynamical chiral symmetry broken is one of the most significant features of QCD, where quark obtains a dynamical mass. It has been confirmed that the AMM  of quarks can also be generated dynamically like dynamical quark mass \cite{Ferrer:2013noa,Chang:2010hb,Ferrer:2008dy} because of spin polarization in the presence of the magnetic field.
	
	The magnetic field also impacts the radiation correction of fermion self-energy, which relates to the AMM coupling between the magnetic field and the fermion spin polarization \cite{Chang:2010hb,Ferrer:2008dy,Schwinger:1948iu,Brekke:1987cc,Bicudo:1998qb,Mohr:2008fa,Ferrer:2015wca}. This generates a new term $\frac{1}{2} q_{f} {\kappa}_{f} {\sigma }_{\mu\nu  } {F}_{\mu\nu }$ in the Dirac Hamiltonian, where $ {\sigma }_{\mu\nu  }=\frac{i}{2} \left [ {\gamma }_{\mu },{\gamma }_{\nu } \right ]$, ${F}_{\mu\nu }$ and ${q}_{f}$ are the spin tensor, field tensor and quark charge, respectively. The AMM term in the Hamiltonian alters the energy spectrum of fermions by eliminating spin degeneracy and influences the properties of the magnetized QCD system \cite{Ferrer:2013noa,Xu:2020yag,Ferrer:2014qka,Mei:2020jzn,Fayazbakhsh:2014mca,Aguirre:2020tiy,Blaizot:2012sd,Ferrer:2009nq}, where the coefficient ${\kappa}_{f}$ is identified as the fermion AMM. Quark AMM reduces the effective quark mass with ${m}_{f}=m-{\kappa}_{f}\left | {q}_{f}B \right | $ in the lowest Landau energy level approximation, and it takes an IMC in the chiral restoration and deconfinement phase transitions.
	
	Modifications related to the QCD phase diagram may also have some impact on the transport characteristics of the QCD medium with the magnetic field generated by relativistic heavy ion collisions (HIC). Different transport coefficients of quark matter have recently been investigated, such as shear viscosity \cite{Huang:2009ue,Huang:2011dc,Tuchin:2011jw,Nam:2013fpa,Alford:2014doa,Tawfik:2016ihn,Hattori:2017qih,Li:2017tgi} and bulk viscosity \cite{Agasian:2011st,Agasian:2013wta} in the presence of the magnetic field. The simulation of magnetic fluid mechanics \cite{Roy:2015kma,Pu:2016ayh} and the transport simulation of the external magnetic field \cite{Das:2016cwd} may need these transport coefficients related to temperature and chemical potential for future study.
	
	The shear coefficient $\eta$  and bulk viscosity coefficient $\zeta$ are essential parameters for quantifying the dissipation process in QCD fluid dynamics evolution under the external magnetic field background. It is particularly noteworthy that $\eta$ and $\zeta$ are very sensitive to the QCD phase transition characteristics in a magnetized media \cite{Sasaki:2008um,Mykhaylova:2020pfk,Soloveva:2021quj}. Of particular interest in QCD are the properties of the transport coefficient near the critical end point, where deconfinement and chiral phase transition begin. Because any changes in shear viscosity and bulk viscosity near ${T}_{c}$ will alter the characteristics of the dynamics evolution of the QCD medium and affect the observations characterizing its expansion kinetics. Obviously, during the development of QCD media, its transport properties are changing as it approaches the phase boundary. Therefore, it is imperative to quantitatively study the relationship between the shear viscosity coefficient near the expected phase transition in the magnetized QCD media and the variation of thermal parameters.
	
	It was found that Refs. \cite{Nam:2013fpa,Alford:2014doa,Tawfik:2016ihn} had not yet explored the components generated by the magnetic fields in the early calculations of the shear viscosity of magnetized materials \cite{Huang:2009ue,Huang:2011dc,Tuchin:2011jw,Nam:2013fpa,Alford:2014doa,Tawfik:2016ihn,Hattori:2017qih,Mohanty:2018eja}.  The decompositions of the components of shear viscosity caused by anisotropy generated by external magnetic fields or other sources in Refs. \cite{Huang:2009ue,Huang:2011dc,Tuchin:2011jw,Mohanty:2018eja} have been well studied in the direction of gauge/gravity duality (see Refs.~\cite{Critelli:2014kra,Jain:2015txa} and their references). The general structure of the shear viscosity coefficients of the five components near the phase boundary is crucial for studying phase transition characteristics.
	
	In this work, we will use the dynamical NJL model \cite{Hatsuda:1994pi,Buballa:2003qv} to study the transport properties of magnetized quark matter near the chiral phase transition. Most of our research on the low energy state of QCD, especially the QCD phase diagram, is based on the quasiparticle model. The condensed state of quarks determines the effective quark mass in the medium. As mentioned above, a strong magnetic field can generate AMM that alters quark condensation. This effect is taken into account for calculating the shear viscosity coefficient in this paper. The results of NJL models with AMM are qualitatively consistent with the lattice simulation results \cite{Aguirre:2020tiy,Blaizot:2012sd,Bali:2012zg}, providing the IMC characteristics of the chiral phase transition, which is also qualitatively consistent with the QCD coupling characteristics. Then, we use the Boltzmann equation's dynamic model and relaxation time approximation to study the features of the viscosity coefficient components in the first-order phase transition and critical end point (CEP) phase transition regions under the strong magnetic field background.
	
	This paper is organized as follows. In Sec. II, we introduce the description of the NJL model with AMM for the magnetized QCD matter. After considering the influence of IMC on thermodynamic physical quantities, we study the chiral phase transition and some thermodynamic properties.  Then, the Boltzmann equation's dynamic model
	and relaxation time approximation are utilized to study the characteristics of the viscosity coefficient components in Sec. III. The characteristics of the shear viscosity coefficient component of a magnetized medium system with first-order phase transitions and critical end point transitions near CEP are investigated in Sec. IV. We summarized and concluded in Sec. V.
	
	\section{THE TWO FLAVORS NJL MODEL UNDER A MAGNETIC FIELD}\label{sec:02 setup}
	
	Let us study the Lagrangian density within a two-flavor NJL model with AMM in the presence of an external magnetic field, which is given by \cite{Fayazbakhsh:2014mca}
	\begin{equation}\label{eq:01}
		\begin{split}
			\mathcal{L}=\bar{\psi }\left( i{{\gamma }^{\mu }}{{D}_{\mu }}+{{\gamma }^{0}}\mu - m +\frac{1}{2}q_{f} \kappa_{f}{\sigma ^{\mu \nu }}{F_{\mu \nu }} \right)\psi + {G}\left[ {{\left( \bar{\psi }{{\lambda }_{a}}\psi  \right)}^{2}}+{{\left( \bar{\psi }i{{\gamma }^{5}}{{\lambda }_{a}}\psi  \right)}^{2}} \right], \\
		\end{split}
	\end{equation}
	where the quark field $\psi $ carries two flavors ($f = u, d$) and three colors ($c=r,g,b$ ). Current quark mass $m$ is considered as ${{m}_{u}}={{m}_{d}}$ for maintaining isospin symmetry, and covariant derivative ${{D}_{u}}={{\partial }_{\mu }}+\operatorname{i}QA_{\mu }^{\operatorname{ext}}$ and charge matrix in flavor space $q_{f} = {\rm{diag}}\left( {{q_u},{q_d}} \right) = {\rm{diag}}\left( {\frac{2}{3}e, - \frac{2}{3}e} \right)$ are introduced by considering the magnetic field. If one selects the gauge field $A_\mu ^{ext} = \left( {0,0,B{x_1},0} \right)$, the corresponding constant magnetic field should point at the $z$ direction. By using mean-field approximation, one can obtain the one-loop effective potential as
	\begin{equation}\label{eq:02}
		\begin{split}
			\Omega  = & \frac{{{\sigma ^2}}}{{4G}} - \frac{{{N_c}\mathop \sum \nolimits_{l = 0}^\infty  {\alpha _l}\mathop \sum \nolimits_{f = u}^d |{q_f}B|}}{{2\pi }}\mathop \int \nolimits_{ - \infty }^\infty  \frac{{d{p_z}}}{{2\pi }} \bigg\{ {\varepsilon _{l,\eta }}+ T\ln \left[ {1 + \exp \left( {\frac{{ - {\varepsilon _{l,\eta }} - \mu }}{T}} \right)} \right] \\
			& +T\ln \left[ {1 + \exp \left( {\frac{{ - {\varepsilon _{l,\eta }} + \mu }}{T}} \right)} \right] \bigg\}, 	
		\end{split}	
	\end{equation}
	where ${E_{_{f,l,t}}} = \sqrt {p_z^2 + {{\left( {{{\left( {{M_f}^2 + 2|{q_f}B|l} \right)}^{1/2}} - \eta {\kappa _f}{q_f}eB} \right)}^2}}$ is the energy spectrum with different Landau energy levels, which $l$ = 0, 1, 2 ... represents the quantum number of Landau level and $\eta  =  \pm 1$ corresponds to the two kinds of spin direction of quark-antiquark pair. Contribution of nondegenerate particles due to spin difference at nonlowest Landau energy levels (non-LLL) should be taken into account with the definition of this operator ${\alpha _l} = {\delta _{0,l}} + \Delta \left( l \right)\sum_{{\eta  =  \pm 1}}^{} $, where the step function $\Delta \left( l \right)$ is taken as $\Delta \left( {l = 0} \right) = 0$ and $\Delta \left( {l > 0} \right) = 1$, respectively.
	One can obtain two coupling gap equations for each order parameter as
	\begin{equation}\label{eq:03}
		\frac{{\partial {\Omega _{{\rm{AMM}}}}}}{{\partial {M_f}}} = 0,
	\end{equation}
	where $f = u,{\rm{ }}d$ for the two different flavors. Then, one can obtain two dynamical quark mass
	\begin{equation}\label{eq:04}
		{M_f} = {m_0} - 2G{\left\langle {\bar \psi \psi } \right\rangle _f}.
	\end{equation}
	
	By quantitatively investigating the contribution of different Landau energy levels to the chiral condensate, We then make the calculations as
	\begin{equation}\label{eq:05}
		{\left\langle {\bar \psi \psi } \right\rangle _f} = \sum\limits_{{\rm{LL = 0}}}^\infty  {{{\left\langle {\bar \psi \psi } \right\rangle }_{{\rm{LL}}}}}  = \sum\limits_{l = 0}^\infty  {\sum\limits_{f = u,d} {{\varphi _{f,l}}} },
	\end{equation}
	where
	\begin{equation}\label{eq:06}
		{\varphi _{f,l}} = \frac{{{N_c}{G_s}}}{{2\pi }}{\alpha _l}|{q_f}B|\int_{ - \infty }^{ + \infty } {{f_\Lambda }(p)\frac{{d{p_z}}}{{2\pi }}\frac{{{M_f}}}{{{\varepsilon _{f,l,t}}}}\left( {1 - \frac{{s{\kappa _f}{q_f}B}}{{{{\hat M}_{f,l,t}}}}} \right)\left\{ {1 - \frac{1}{{{e^{\frac{{{\varepsilon _{f,l,t}} + \mu }}{T}}} + 1}} - \frac{1}{{{e^{\frac{{{\varepsilon _{f,l,t}} - \mu }}{T}}} + 1}}} \right\}},
	\end{equation}
	corresponds to chiral condensation of different quark flavors ($f = u,d$). As is well known, the NJL model is nonrenormalized, so the regularization scheme is necessary for extracting finite numerical results. A soft truncation by introducing a smooth truncation function ${f_\Lambda }(p)$ ~\cite{Ferrer:2015wca} is adopted to ensure its convergence
	during the momentum integration as
	\begin{equation}\label{eq:07}
		{f_\Lambda }(p) = {\left( {1 + \exp \left( {\frac{{\sqrt {{p^2} + 2l\left| {{q_f}eB} \right|}  - \Lambda }}{{0.05\Lambda }}} \right)} \right)^{ - 1}}.
	\end{equation}
	
	\begin{figure}[H]
		\centering
		\includegraphics[width=0.35\textwidth]{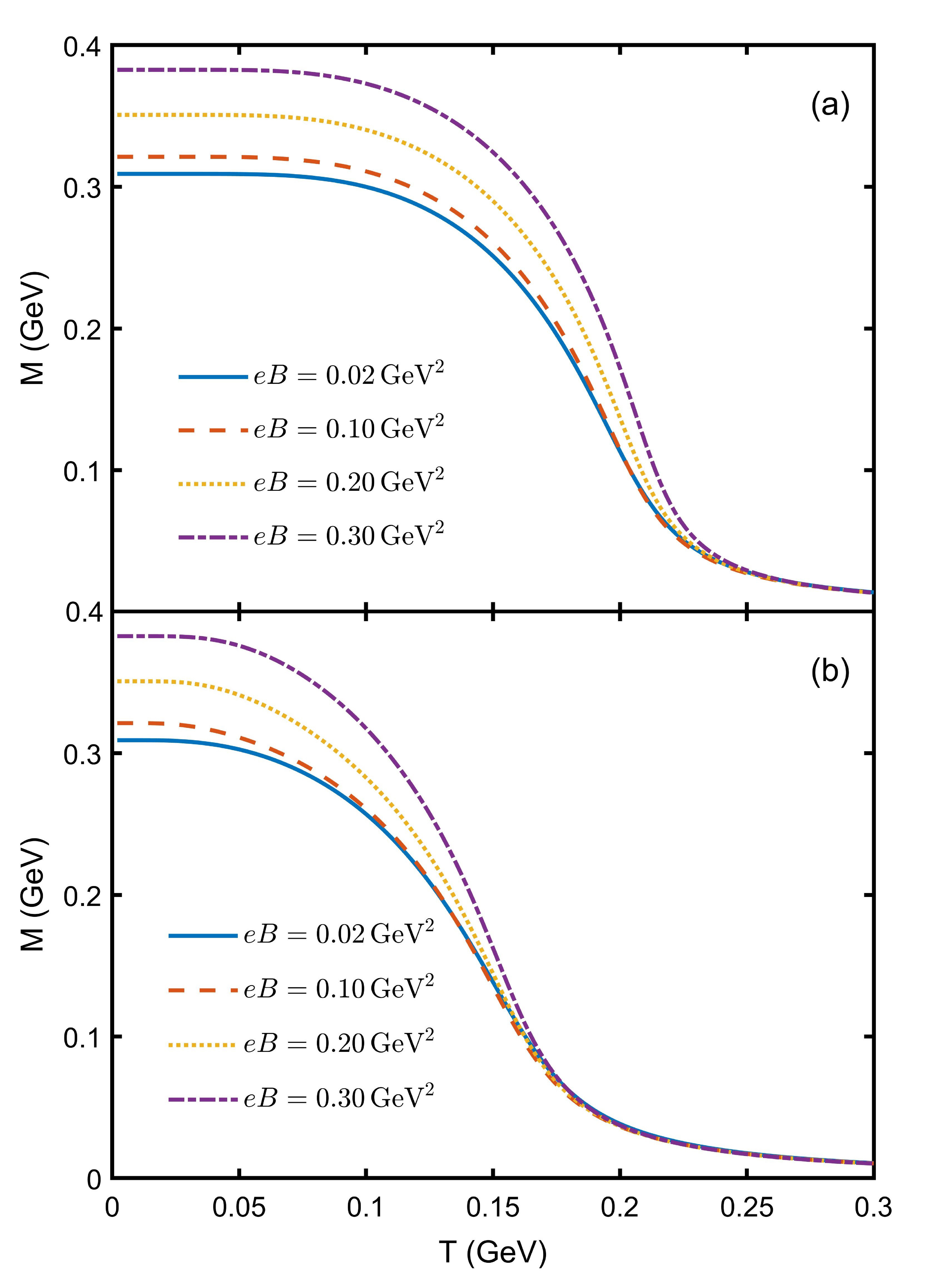}
		\caption{\label{fig1} The dependence of dynamical quark mass ($M$) on temperature (T) for four different magnetic fields ($eB = $0.02, 0.10, 0.20, and 0.30~${\rm{Ge}}{{\rm{V}}^{\rm{2}}}$) for standard NJL model without AMM. (a) is for $\mu  = $0.00 GeV, and (b) is for $\mu  = $0.225 GeV.}
	\end{figure}
	It is well known that the dynamical mass or the quark condensate acts as an order parameter for the chiral-phase transition. Chiral restoration occurs at high temperatures and high chemical potentials. In Figs. 1(a) and 1(b), the dynamical quark mass ${M}$ of $u$ and $d$ quarks without considering AMM are manifested as decreasing smooth functions of temperatures at $\mu  = $ 0 GeV and $\mu  = $ 0.225 GeV, which indicates a chiral crossover. Since we have considered the nonvanishing current quark mass, the chiral symmetry is never restored fully. Since the dynamical mass is proportional to chiral condensate, it can be seen from Fig. 1 that the larger the magnetic field is, the larger the corresponding chiral condensation is. This phenomenon is manifested as magnetic catalysis \cite{Kharzeev:2007jp,Gusynin:1999pq,Gusynin:1995nb,Gusynin:1994re}, which accounts for the magnetic field having a solid trend for lift (or catalyze) spin-zero $q\bar q$ condensates.
	
	Figure 2 displays the dependence of dynamical quark mass (${M}$) on temperature (T) for four different magnetic fields ($eB = $ 0.02, 0.10, 0.20, and 0.30~${\rm{Ge}}{{\rm{V}}^{\rm{2}}}$) by considering the AMM. Figures 2(a) and 2(b) are for $\mu  = $ 0.00 GeV and 0.225 GeV, respectively. Contrary to the behavior of the without AMM shown in Fig. 1, the dynamical mass-decreasing behavior of $u$ and $d$ quarks in the chiral restoration is not a smooth function but a sudden drop, which indicates the existence of a first-order transition. However, the slippery slope of the dynamical mass for the crossover can still be present in the weak field $eB = $ 0.02 ${\rm{Ge}}{{\rm{V}}^{\rm{2}}}$ . It is found that the dynamical quark mass of $u$ and $d$ quarks have the characteristics of IMC in the chiral restoration phase ($T \ge {T_C}$) by inducing AMM.
	\begin{figure}[H]
		\centering
		\includegraphics[width=0.35\textwidth]{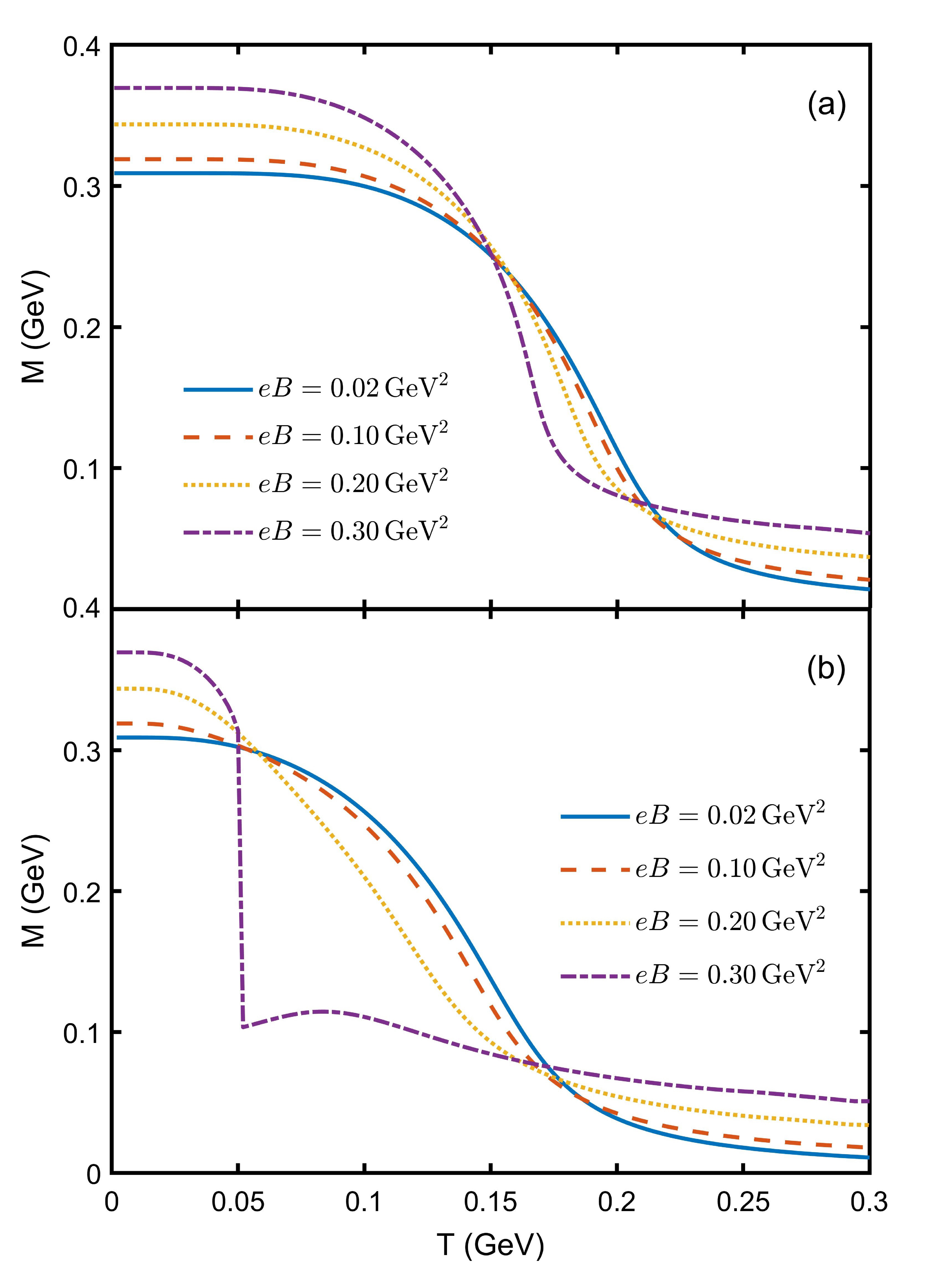}
		\caption{\label{fig2} The dependence of dynamical quark mass ($M$) on temperature (T) for four different magnetic fields ($eB = $ 0.02, 0.10, 0.20, and 0.30~${\rm{Ge}}{{\rm{V}}^{\rm{2}}}$) with AMM. (a) is for $\mu  = $ 0.00 GeV, and (b) is for $\mu  = $ 0.225 GeV.}
	\end{figure}
	
	The contribution ratios from different Landau levels to chiral condensation under strong magnetic fields are manifested in Fig. 3. The contribution ratios for considering AMM shown in Figs. 3(a) and 3(b) are compared to the ratios for without considering AMM shown in Figs. 3(c) and 3(d).  It is found that the contribution ratio of each Landau level of the NJL model without inducing AMM to the chiral condensation does not change significantly with increasing temperature. Even when the temperature rises to the chiral phase transition temperature where the chiral symmetry is restored, the contribution of each different Landau level to the chiral condensation is roughly unchanged.
	\begin{figure}[H]
		\centering
		\includegraphics[width=0.65\textwidth]{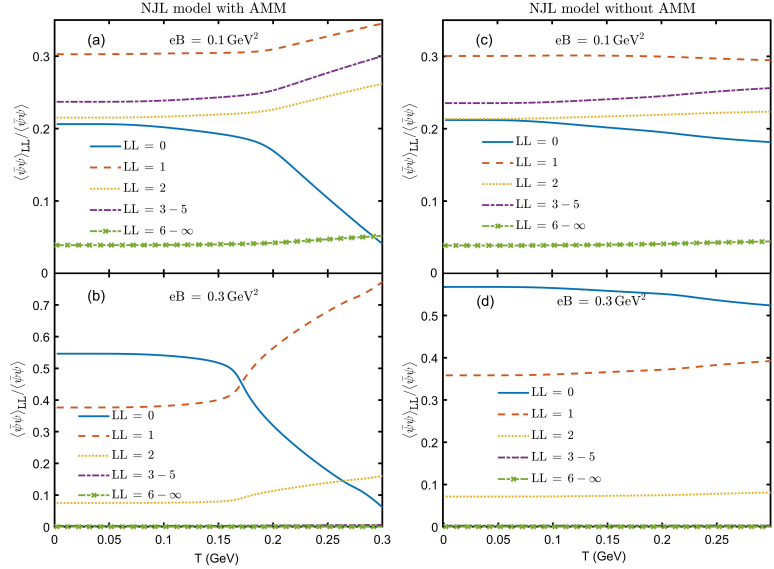}
		\caption{\label{fig3} The temperature dependence of the contribution ratio of quark condensation with different Landau levels under various magnetic fields in zero chemical potential. Figures 3(a) and 3(b) by considering AMM are for $eB = $ 0.1 ${\rm{Ge}}{{\rm{V}}^{\rm{2}}}$ and $eB = $ 0.3 ${\rm{Ge}}{{\rm{V}}^{\rm{2}}}$, respectively, while Figs.3(c) and 3(d) are the same as Figs. 3(a) and 3(b), but without considering AMM.}
	\end{figure}
	As follows, we will focus on studying the contribution ratio of the NJL model by inducing AMM. It is found that as the temperature increases, after reaching the chiral phase transition temperature, the contribution ratio of the lowest Landau level to chiral condensation decreases significantly with temperature. Still, the contribution ratio of the excited Landau level increases with temperature accordingly, as shown in Figs. 3(a) and 3(b). From the analysis, we can understand that the effect of AMM can make particles with the lowest Landau level more easily excited to higher excited energy levels at high temperatures. And as the magnetic field increases, this effect becomes more pronounced. It is believed that this may be the reason for the formation of IMC at high temperatures after the introduction of AMM.
	
	The dependence of the renormalized critical temperature $T_{c}/T_{0}$ on the magnetic field $eB$ is shown in Fig. 4.  $T_{0}$ is the critical temperature when the magnetic field is zero. It can be seen that the NJL model with AMM has a noticeable IMC effect. That is, with the increase of the magnetic field, the critical temperature $T_c$ gradually decreases with magnetic field. On the other hand, the NJL model without AMM, exhibits an MC characteristic, where the critical temperature increases with the increase of the magnetic field. Our explanation for this phenomenon is that the dimensional reduction in the magnetic field background causes particles to occupy different excited Landau levels. Although the contribution of Landau levels in the low-temperature region is less affected by temperature, the effect of the inducing AMM contributes to higher Landau levels becoming great near the critical temperature, as shown in Fig. 3.  Without considering AMM, it would have led us to overestimate the effective kinetic mass of quarks, resulting in the MC effect. At high temperatures, the impact of more particles exciting to higher Landau levels will counteract some of the MC effects and display IMC.
	\begin{figure}[H]
		\centering
		\includegraphics[width=0.45\textwidth]{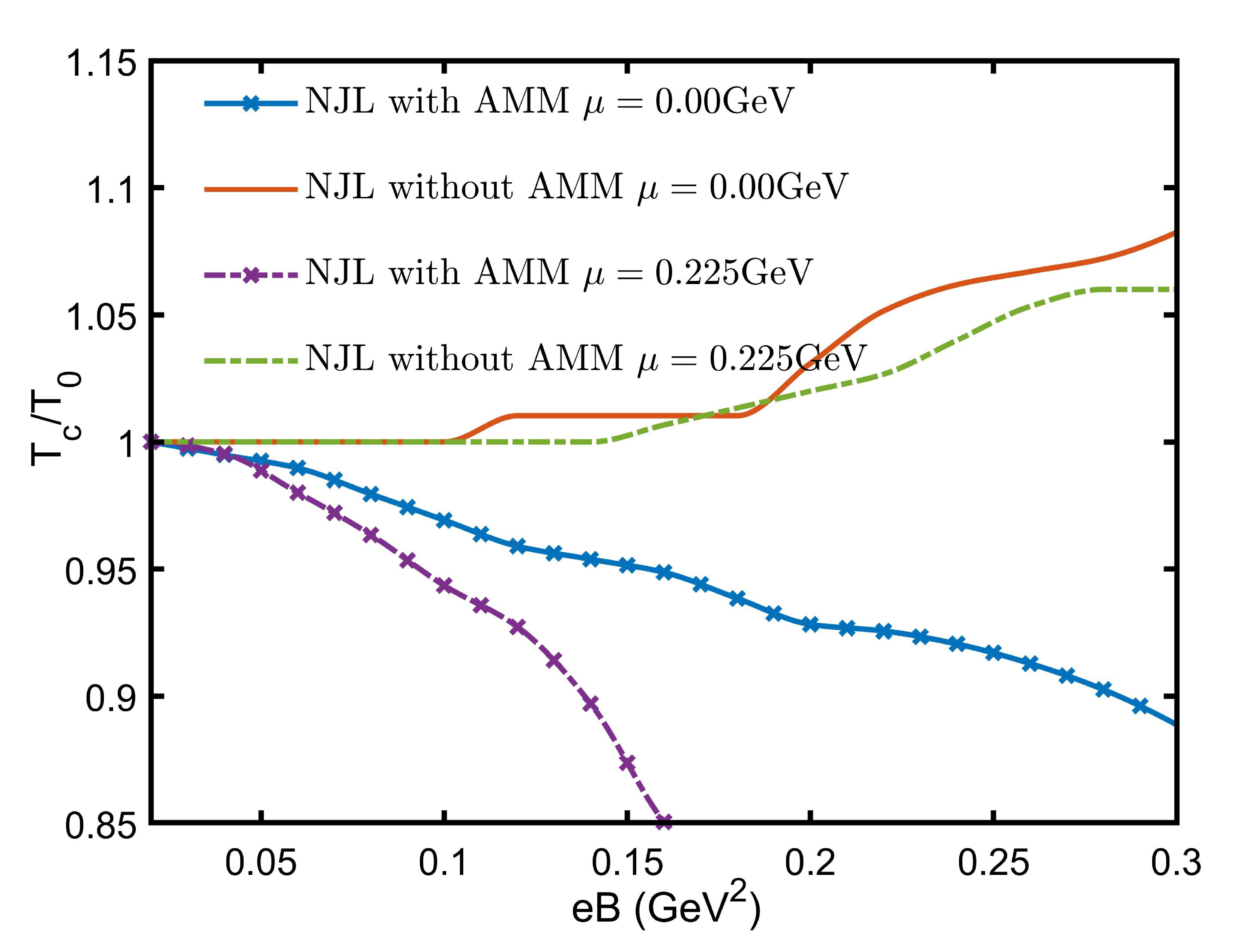}
		\caption{\label{fig4} The dependence of renormalized critical temperature on magnetic
			field for NJL model with or without the consideration of AMM at zero and finite
			chemical potentials.}
	\end{figure}
	
	\begin{figure}[H]
		\centering
		\includegraphics[width=0.45\textwidth]{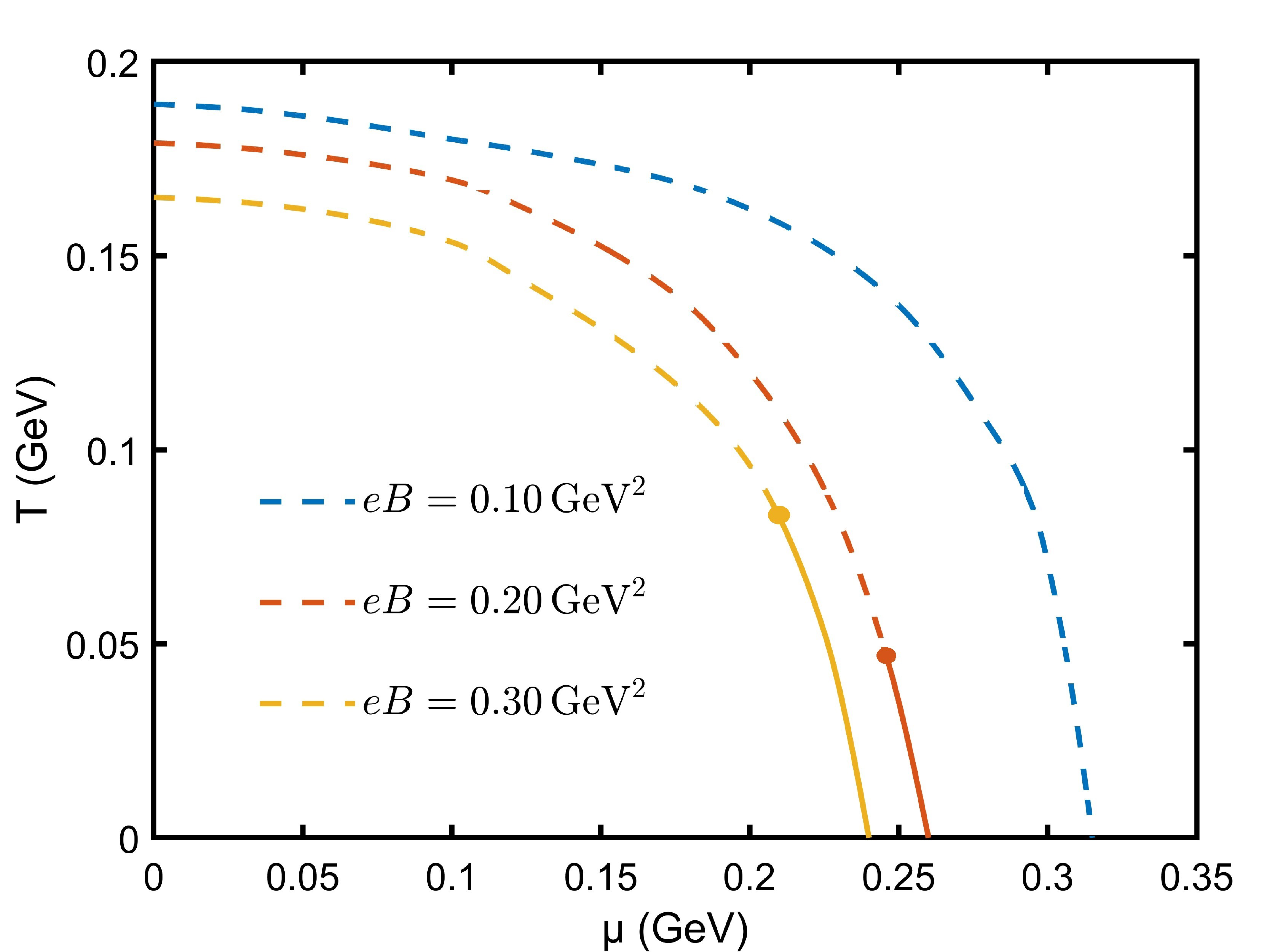}
		\caption{\label{fig5} Phase diagrams in the presence of strong magnetic fields $eB = $ 0.1, 0.2, and 0.3 ${\rm{Ge}}{{\rm{V}}^{\rm{2}}}$, respectively. The solid lines correspond to first-order transition, the dashed lines correspond to crossover phase transition, and the full dots correspond to CEP.}
	\end{figure}
	The phase diagrams by introducing AMM with magnetic fields of $eB$ = 0.10, 0.2, and $eB$ = 0.30 ${\rm{Ge}}{{\rm{V}}^{\rm{2}}}$ are published in Fig. 5. It is found that when the magnetic field is as low as  $eB$ = 0.1 ${\rm{Ge}}{{\rm{V}}^{\rm{2}}}$, the phase transition is crossover transition, and no first-order phase transition occurs. When the magnetic field rises from $eB=$ 0.2 to 0.3 ${\rm{Ge}}{{\rm{V}}^{\rm{2}}}$, the crossover occurs at high temperatures and small chemical potentials $\mu$, while the first-order phase transition happens at low temperatures $T$ and large chemical potential $\mu$. The results are consistent with those of Refs. \cite{Zhu:2022irg,Su:2021cda}.
	The phase transition between the crossover and first order transition is separated by the critical end point (CEP) in $T-\mu$ plane. The existence of CEP is an important topic in the QCD phase diagram, and lattice QCD and some effective models have investigated the location of CEP \cite{Zhu:2022irg,Fodor:2004nz,Costa:2015bza}. Figure 5 displays the locations of CEP for $eB$ = 0.2 ${\rm{Ge}}{{\rm{V}}^{\rm{2}}}$ and $eB$ = 0.3 ${\rm{Ge}}{{\rm{V}}^{\rm{2}}}$, respectively. When $eB$ = 0.2 ${\rm{Ge}}{{\rm{V}}^{\rm{2}}}$ and $eB$ = 0.3 ${\rm{Ge}}{{\rm{V}}^{\rm{2}}}$, the CEP are located at ($T$, $\mu$) = (47.3, 245.7) MeV and ($T$, $\mu$) = (76.1, 209.7) MeV, respectively. The temperature of CEP will increase with the magnetic field, while the chemical potential of the CEP decreases with the temperature of CEP.

	\section{Boltzmann equation in relaxation time approximation and shear viscosity coefficient}\label{sec:03 setup}
	The produced medium can be supposed to be a kind of near equilibrium thermal system. Therefore, we can investigate the transport properties of the nearly perfect fluid impacted by QCD phase transition. The shear viscosity of the medium can be defined by the Boltzmann equation of the relativistic fluid, and the relaxation time of quarks can be calculated from the thermally averaged total cross section of elastic scattering. The local equilibrium distribution function for the quarks is given as
	\begin{equation}\label{eq:08}
		f(x,p) = {f^0}(x,p) + {f^1}(x,p),
	\end{equation}
	where ${f^0}(x,p)$ is the equilibrium distribution function as
	\begin{equation}\label{eq:09}
		{f^0}(x,p) = {\left[ {\exp \left( {\beta \left( x \right)\left( {{u_\mu }\left( x \right){p^\mu } \mp \mu \left( x \right)} \right)} \right) + 1} \right]^{ - 1}},
	\end{equation}
	where ${u_\mu }\left( x \right)$ is the four-velocity of the fluid, and $\mu \left( x \right)$ is the chemical potential. The covariant Boltzmann equation of motion can be taken as
	\begin{equation}\label{eq:10}
		\frac{{df}}{{dt}} = \frac{{{p^\mu }}}{E}{\partial _\mu }f - \frac{{\partial E}}{{\partial {x^i}}}\frac{{\partial f}}{{\partial {p^i}}} =  - \frac{{{f^1}}}{\tau },
	\end{equation}
	where $\tau$ is the relaxation time. The total energy-momentum tensor of the fluid is given as ${T_{\mu \nu }} = T_{\mu \nu }^0 + T_{\mu \nu }^D$, where the $T_{\mu \nu }^0 =  - p{g^{\mu \nu }} + w{u^\mu }{u^\nu }$ is the ideal part, and the dissipative part is given as
	\begin{equation}\label{eq:11}
		T_{\mu \nu }^D = \eta \left( {{D^\mu }{u^\nu } + {D^\nu }{u^\mu } + \frac{2}{3}{\Delta ^{\mu \nu }}{\partial _\alpha }{u^\alpha }} \right) - \zeta {\kern 1pt} {\partial _\alpha }{u^\alpha },
	\end{equation}
	where $\eta$ is the shear viscosity coefficient and $\zeta$ is the bulk viscosity coefficient. The total energy-momentum tensor can be expressed as
	\begin{equation}\label{eq:12}
		{T_{\mu \nu }} = \int {\frac{{{d^3}k}}{{{{(2\pi )}^3}}}} \frac{{{k^\mu }{k^\nu }}}{\omega }\left( {{f^0} + \Phi {\kern 1pt} {f^0}\left( {1 - {f^0}} \right)} \right).
	\end{equation}
	
	The shear viscosity coefficient $\eta$ can be obtained by comparing the dissipative part of the energy-momentum tensor and possess five independent components in the presence of an external magnetic field \cite{Tuchin:2011jw},
	\begin{equation}\label{eq:13}
		\eta = \frac{1}{{15T}}\sum\limits_a {{g_a}} \int {\frac{{{d^3}{{\vec k}_a}}}{{{{(2\pi )}^3}}}} \frac{{{{\vec k}_a}^4}}{{{E_a}^2}}\tau _{a}^{c}\left( {f_a^0\left( {1 - f_a^0} \right)} \right),
	\end{equation}
	wherein the viscosity coefficient of ${\eta}_{0}$ is the same as the viscosity coefficient when the magnetic field is zero.
	These remaining four components can be written as
	\begin{equation}\label{eq:14}
		{\eta _{n(n = 1,2,3,4)}} = \frac{1}{{15T}}\sum\limits_a {{g_a}} \int {\frac{{{d^3}{{\vec k}_a}}}{{{{(2\pi )}^3}}}} \frac{{{{\vec k}_a}^4}}{{{E_a}^2}}\tau _{a,n}^{eff}\left( {f_a^0\left( {1 - f_a^0} \right)} \right),
	\end{equation}
	where ${g_a}$ is the degeneration of collisional particle type $a$, the  $\tau _{a,n}^{eff}$ is the effective relaxation time, which $\tau _a^B$ is related to the magnetic relaxation time and collision relaxation time $\tau _a^c$. The approximation of the field-induced relaxation time $\tau _a^B = E{}_a/({q_a}B)$ has been applied in the strong magnetic field since the deviation of equilibrium state mostly results from the field rather than the particle collisions. The  ansatz of effective relaxations $\tau _{a,n}^{eff}$~~\cite{Ghosh:2018cxb,Critelli:2014kra} is taken as
	\begin{equation}\label{eq:15}
		\begin{array}{l}
			\tau _{a,1}^{eff} = \tau _a^c\frac{1}{{4\{ \frac{1}{4} + {{(\tau _a^c/\tau _a^B)}^2}\} }},\quad \tau _{a,2}^{eff} = \tau _a^c\frac{1}{{\{ 1 + {{(\tau _a^c/\tau _a^B)}^2}\} }}\\
			\tau _{a,3}^{eff} = \tau _a^c\frac{{\tau _a^c/\tau _a^B}}{{2\{ \frac{1}{4} + {{(\tau _a^c/\tau _a^B)}^2}\} }},\quad \tau _{a,4}^{eff} = \tau _a^c\frac{{\tau _a^c/\tau _a^B}}{{\{ 1 + {{(\tau _a^c/\tau _a^B)}^2}\} }}.
		\end{array}
	\end{equation}
	
	By considering the elastic scattering of two-body for this two-flavor system, one can obtain 12 different collision processes. In order to determine the collision relaxation time of the system, the elastic scattering theory of the two-body collision $a + b \to c + d$ is utilized in our study. The constituent particles of the system medium are mainly $u$ and $d$ quarks and their antiquarks near the phase transition temperature. A total of 12 collision scenarios are taken as
	\begin{equation}\label{eq:16}
		\begin{array}{*{20}{c}}
			{uu \to uu,}&{ud \to ud,}&{u\bar u \to u\bar u,}&{u\bar u \to d\bar d,}\\
			{u\bar d \to u\bar d,}&{dd \to dd,}&{\bar ud \to \bar ud,}&{d\bar d \to d\bar d,}\\
			{d\bar d \to u\bar u,}&{\bar u\bar u \to \bar u\bar u,}&{\bar u\bar d \to \bar u\bar d,}&{\bar d\bar d \to \bar d\bar d.}
		\end{array}
	\end{equation}
	
	One assumes that particle $a$ is a probe particle, and the collision relaxation time of the system could be evaluated $\tau _a^c = \frac{1}{{\Gamma _a^c}}$. The collision width $\Gamma _a^c$ is given as
	\begin{equation}\label{eq:17}
		\begin{array}{*{20}{c}}
			\Gamma _a^c = \frac{{\int {\frac{{{d^3}{p_a}}}{{{{\left( {2\pi } \right)}^3}}}\Gamma _a^c({p_a})} {f_a}}}{{\int {\frac{{{d^3}{p_a}}}{{{{\left( {2\pi } \right)}^3}}}} {f_a}}},
		\end{array}
	\end{equation}
	where the $\Gamma _a^c({p_a})$  is taken as
	\begin{equation}\label{eq:18}
		\Gamma _a^c({p_a}) = \sum\limits_a {\int {\frac{{{d^3}{p_b}}}{{{{\left( {2\pi } \right)}^3}}}{\sigma _{ab}}\left( {{p_a},{p_b}} \right)} } {\nu _{ab}}\left( {{p_a},{p_b}} \right){f_b}.
	\end{equation}
	where ${\sigma _{ab}}$ and ${\nu _{ab}}$ are the collision cross section and the relative velocity, respectively, which are given as
	\begin{equation}\label{eq:19}
		\frac{d\sigma _{ab}}{dt} = \frac{1}{{16\pi s}} \frac{1}{p_{ab}^{2}}{\left| {{{\tilde M}_{ab}}}\right|^2},
	\end{equation}
	and
	\begin{equation}\label{eq:20}
		{\nu _{ab}}\left( {{p_a},{p_b}} \right) = \frac{{\left( {{E_a} + {E_b}} \right)\sqrt {{s^2} - 4{M^2}} }}{{2{E_a}{E_b}}},
	\end{equation}
	where $s = {\left( {{E_a} + {E_b}} \right)^2}$ is the Mandelstam variable, and the scattering matrix ${\left| {{{\tilde M}_{ab}}} \right|^2}$ in Eq. (19) can be simplified from 12 situations to only two independent cases by the isospin symmetry, charge conjugation and crossing symmetries. The results  \cite{Zhuang:1995uf} for the $u\bar{u}\rightarrow u\bar{u}$ process, and the $u\bar{d}\rightarrow u\bar{d}$ case can be given as
	\begin{equation}\label{eq:21}
		\begin{split}
			{\left| {{{\tilde M}_{u\bar{u}\rightarrow u\bar{u}}}} \right|^2} =&{s^2}\left|D_\pi\left(\sqrt{s},0\right)\right|^{2}+{t^2}\left|D_\pi\left(0,\sqrt{-t}\right)\right|^{2}\left(s-4m^{2}\right)^{2}\left|D_\sigma\left(\sqrt{s},0\right)\right|^{2}+\left(t-4m^{2}\right)^{2}\left|D_\sigma\left(0,\sqrt{-t}\right)\right|^{2}\\
			+&\frac{1}{N_{c}}\text{Re}\big[stD_\pi^*\left(\sqrt{s},0\right)D_\pi\left(0,\sqrt{-t}\right)+s\left(4m^{2}-t\right)D_\pi^*\left(\sqrt{s},0\right)D_\sigma\left(0,\sqrt{-t}\right)\\
			&+t\left(4m^{2}-s\right)D_\pi\left(0,\sqrt{-t}\right)D_\sigma^*\left(\sqrt{s},0\right)+\left(4m^{2}-s\right)\left(4m^{2}-t\right)D_\sigma^*\left(\sqrt{s},0\right)D_\sigma^*\left(0,\sqrt{-t}\right) \big].
		\end{split}
	\end{equation}
	and
	\begin{equation}\label{eq:22}
		\begin{split}
			{\left| {{{\tilde M}_{u\bar{d}\rightarrow u\bar{d}}}} \right|^2} =
			&4{s^2}\left|D_\pi\left(\sqrt{s},0\right)\right|^{2}+{t^2}\left|D_\pi\left(0,\sqrt{-t}\right)\right|^{2}\left(s-4m^{2}\right)^{2}\left|D_\sigma\left(\sqrt{s},0\right)\right|^{2}+\left(t-4m^{2}\right)^{2}\left|D_\sigma\left(0,\sqrt{-t}\right)\right|^{2}\\
			+&\frac{1}{N_{c}}\text{Re}\big[-2stD_\pi^*\left(\sqrt{s},0\right)D_\pi\left(0,\sqrt{-t}\right)+2s\left(4m^{2}-t\right)D_\pi^*\left(\sqrt{s},0\right)D_\sigma\left(0,\sqrt{-t}\right)\big],
		\end{split}
	\end{equation}
	respectively. In the random phase approximation (PRA), the effective meson propagator $D_M$ above can be expressed as
	\begin{equation}\label{eq:23}
		D_M\left(\omega,\mathbf{p}\right)=\frac{2iG}{1-2G\Pi_M\left(\omega,\mathbf{p}\right)},
	\end{equation}
the above $M=\pi$ or $\sigma$ meson, and the $\Pi_M$ is the polarization function corresponding the $M$ mesonic channel with AMM under the magnetic field \cite{Xu:2020yag}. The matrix elements on the lowest order $1/N_c$ within the NJL model expressed in Eq. (21) and Eq. (22) consider the contribution of $\pi$ and $\sigma$ meson exchange in $s$ and $t$ channels during the 2$\rightarrow$2 scattering process. The detailed calculation process of the total cross section can be found in Ref.~\cite{Zhuang:1995uf}.
	
	The dependences of the collision relaxation time ${\tau _c}$ on temperature with different magnetic fields with AMM Figs.6(a) and 6(b) and without AMM Figs.6(c) and 6(d) at chemical potential $\mu$ = 0 [Figs. 6(a) and 6(c)] and $\mu$ = 0.225 GeV [Figs. 6(b) and 6(d)] are, respectively, displayed in Fig. 6. We compared the relaxation time with and without the introduction of AMM. From the curve of relaxation time changing with temperature, it can be seen that the relaxation time decreases rapidly with increasing temperature at low temperatures. The increased behaviors of relaxation time with temperature occur in the later stage of the process of restoring chiral symmetry; it still can be observed in the enlarged picture in Fig. 6 though they are very slight. It is found that adding AMM will make the minimum point more obvious and the increased amplitude more significant under high temperature conditions.
	
It is observed that when the temperature is near the phase transition temperature $160-220$ MeV at $\mu$ = 0 with AMM, the corresponding relaxation time is below 10 fm. The relaxation time decreases faster with temperature when the chemical potential $\mu$ = 0.225 GeV than that of the zero-chemical potential, especially it almost drops to zero when the magnetic field becomes large as $eB$ = 0.3 ${\rm{Ge}}{{\rm{V}}^{\rm{2}}}$. We also find that the relaxation time, both of zero and finite chemical potential, have a slow rebound with increasing temperature when the temperature exceeds 0.20 GeV. In the following, we will use Eq. (14) to study shear viscosities near the chiral phase transition of the magnetized QCD medium.
	\begin{figure}[H]
		\centering
		\includegraphics[width=0.65\textwidth]{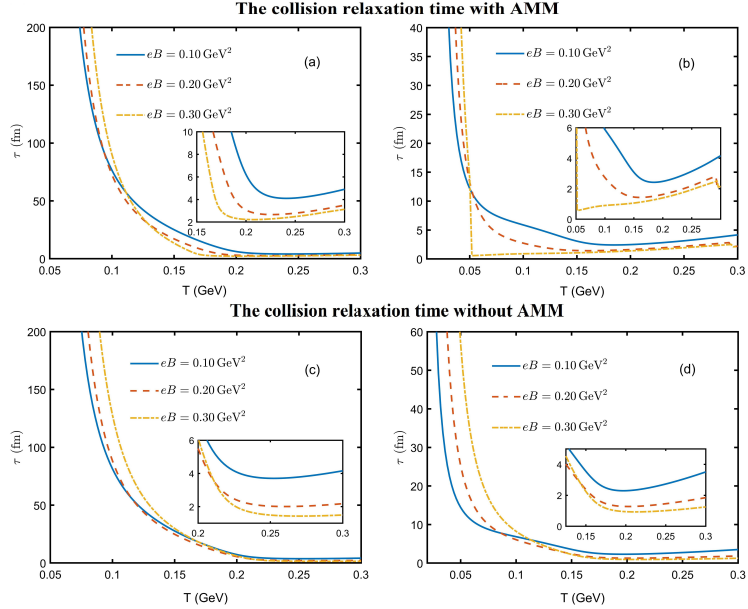}
		\caption{\label{fig6} The temperature dependencies of collision relaxation time with different magnetic fields. Figures 6(a) and 6(b) is for relaxation time by considering AMM, and 6(c) and 6(d) is for relaxation time without considering AMM. Figures 6(a) and 6(c) is for zero chemical potential, and 6(b) and 6(d) is for the finite chemical potential $\mu$ = 0.225 GeV.}
	\end{figure}
	\section{The calculated results of the shear viscosities near chiral phase transition}\label{sec:04 setup}
	The temperature dependence of the ratio $\eta /s$  of shear viscosity coefficient to entropy density under different magnetic fields $eB$ = 0.1, 0.2, and 0.3 ${\rm{Ge}}{{\rm{V}}^{\rm{2}}}$ at $\mu$ = 0 [Fig.7(a)] and $\mu$ = 0.225 GeV [Fig.7(b)] are, respectively, displayed in Fig. 7. It is found that the ratio $\eta /s$ decreases with increasing temperature at low temperature region, and exhibits an upward turn around the critical temperature and finally, begins to increase with temperature. The trend of both $\eta /s$ ratio and relaxation time $\tau$ with temperature have some similar characteristics. Figure 6 shows that at low temperatures, the collision relaxation time $\tau$ rapidly decreases with temperature, but have a little rebound around the critical temperature, and then slightly increases in the high temperature region. It can be clearly observed from Eq.(13) that the collision relaxation time plays a decisive role for the the ratio $\eta /s$ calculation. There have been many studies on the ratio $\eta /s$ of shear viscosity to entropy density~\cite{Csernai:2006zz,Kovtun:2004de}, and Ref.~\cite{Csernai:2006zz} found that for all known substances, this ratio $\eta /s$ reaches a minimum near the critical point. The absolute minimum limit of $\eta /s$ estimated within the AdS/CFT correspondence~\cite{Kovtun:2004de} is $1/(4\pi)$.

	\begin{figure}[H]
		\centering
		\includegraphics[width=0.38\textwidth]{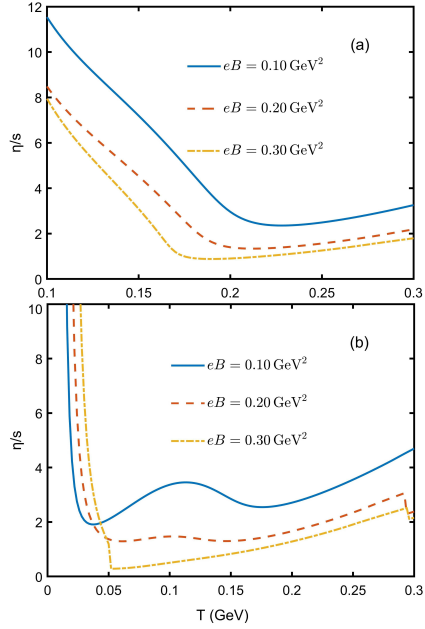}
		\caption{\label{fig7} The temperature dependence of the ratio $\eta/s$ of shear viscosity coefficient to entropy density under different magnetic fields $eB$ = 0.1, 0.2, and 0.3 ${\rm{Ge}}{{\rm{V}}^{\rm{2}}}$. (a) is for zero chemical potential, and (b) is for the finite chemical potential $\mu$ = 0. 225 GeV.}
	\end{figure}
	\begin{figure}[H]
		\centering
		\includegraphics[width=0.42\textwidth]{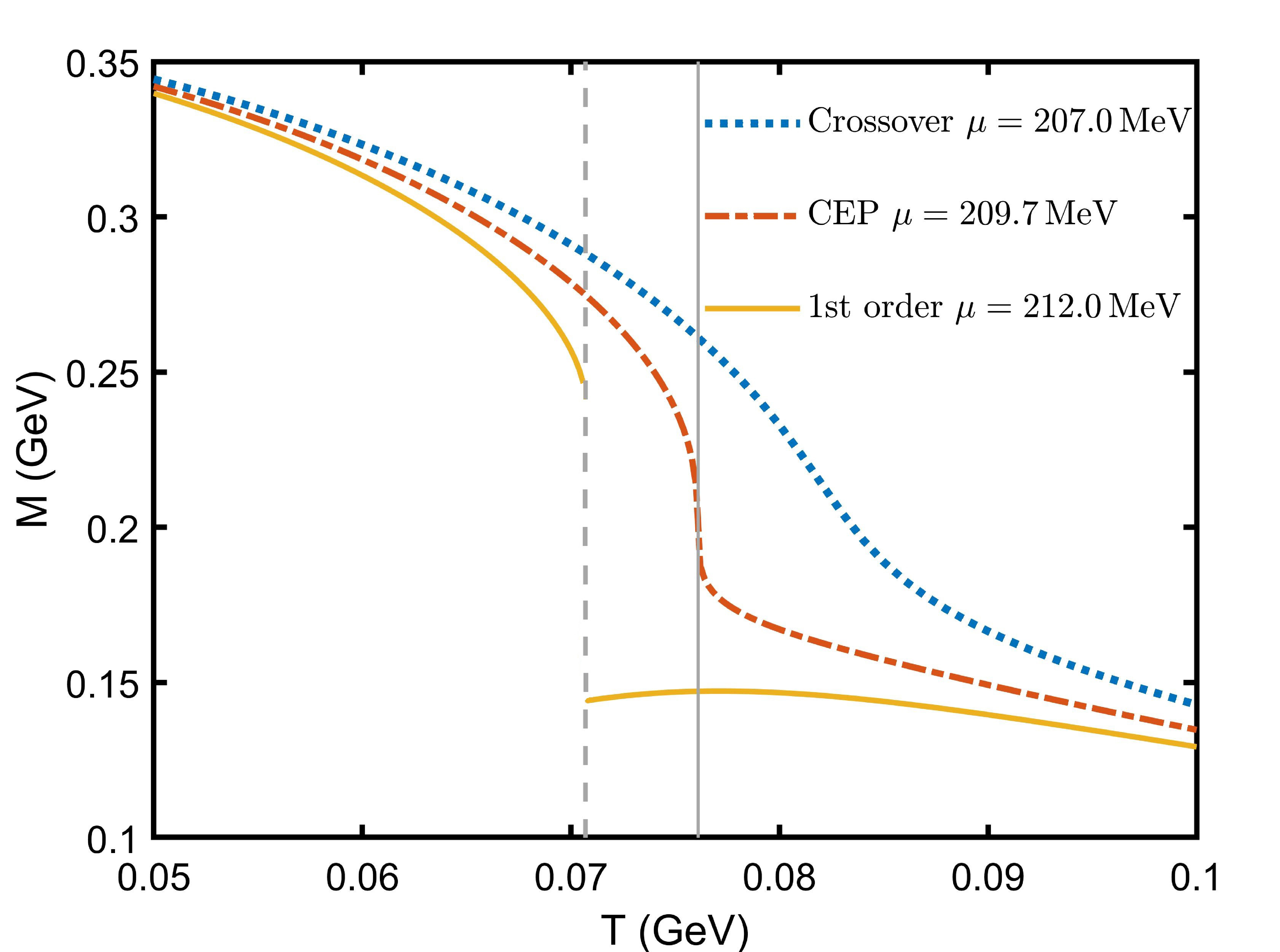}
		\caption{\label{fig8} The dependencies of the dynamic quark mass $M$ on the temperature of the crossover, the first order phase transition and the CEP transition, respectively. The yellow solid line corresponds to the first$-$order phase transition, the red dashed$-$line corresponds to the CEP phase transition, and the blue dotted line corresponds to the crossover phase transition, respectively.}
	\end{figure}
	Figure 8 shows the dependence of the dynamic quark mass $M$ on the temperature of the crossover phase transition, the first order phase transition and the critical end point (CEP) phase transition, respectively. For the crossover phase transition, the thermodynamic potential of the system has only one extreme point, which corresponds to the fact that the dynamic mass varies continuously with temperature. The first-order phase transition, which refers to the phase transition from the hadron phase to the quark-gluon plasma phase, has clear phase boundaries. For the first order phase transition, its dynamic mass $M$ has a discontinuous change at the phase transition point, which is called an abrupt change. For a first-order phase transition, the dynamic mass $M$ undergoes a discontinuous change at the phase transition point, which is called a sudden change. The so-called CEP phase transition is the connection point between the first-order phase transition and the crossover phase transition on the phase diagram and also the end point of the first-order phase transition.
	\begin{figure}[H]
		\centering
		\includegraphics[width=0.42\textwidth]{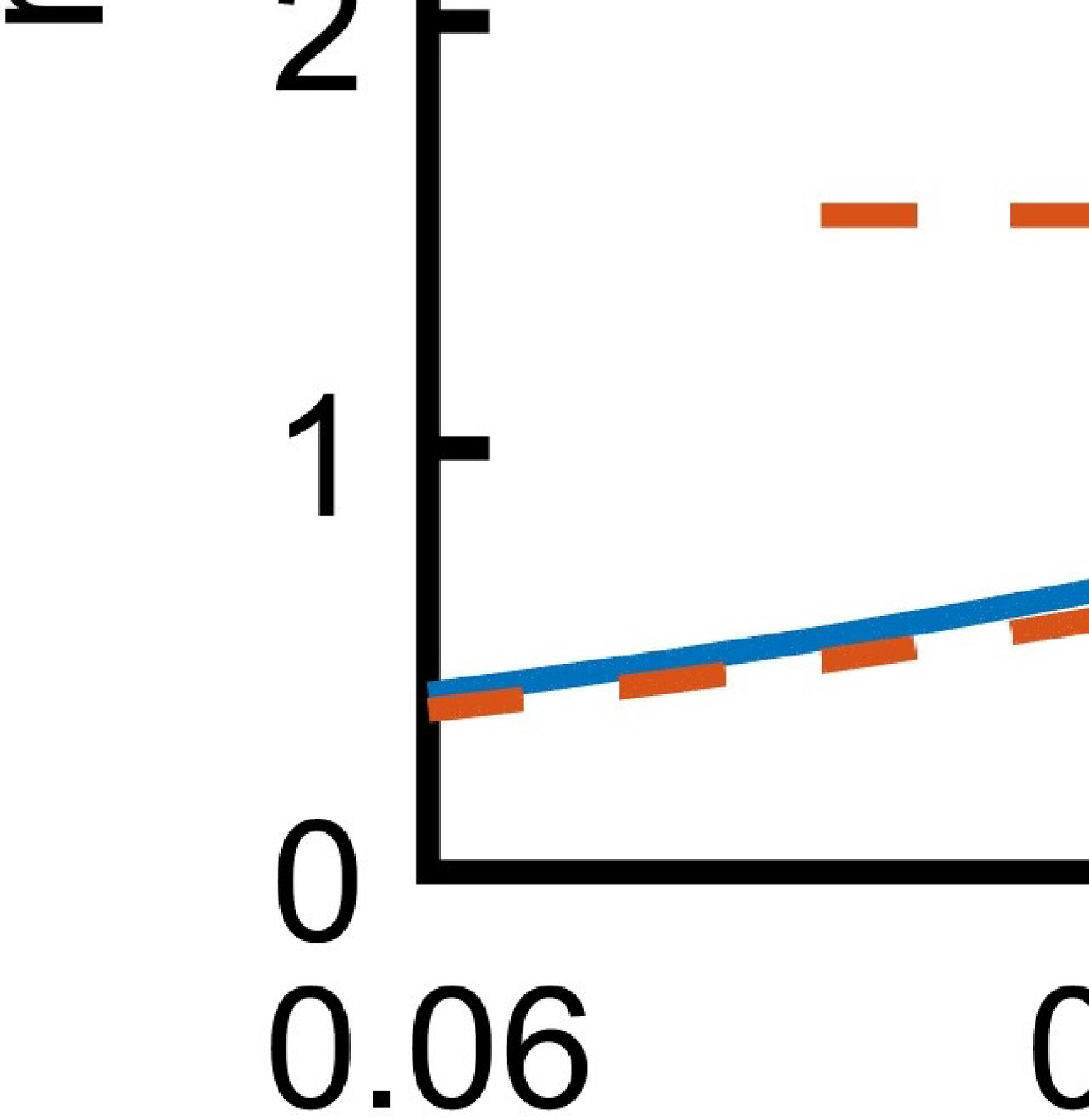}
		\caption{\label{fig9} The dependencies of renormalized shear viscosity coefficient components (${\eta _{1,2,3,4}}/{\Lambda ^3}$) on the temperature of CEP phase transition and first order phase transition respectively with magnetic field ($eB$ = 0.2 ${\rm{Ge}}{{\rm{V}}^{\rm{2}}}$).  (a) for $\eta_{1}/\Lambda^{3}$, (b) for $\eta_{2}/\Lambda^{3}$, (c) for $\eta_{3}/\Lambda^{3}$ and (d) for $\eta_{4}/\Lambda^{3}$.}
	\end{figure}
	The dependencies of renormalized shear viscosity coefficient components (${\eta _{1,2,3,4}}/{\Lambda ^3}$) on temperature of CEP phase transition and first order phase transition with $eB$ = 0.2 ${\rm{Ge}}{{\rm{V}}^{\rm{2}}}$ are shown in Fig. 9(a, b, c and d).  The chemical potential $\mu  = {\mu _{{\rm{CEP}}}}$ is fixed to study the shear viscosity coefficient components for the CEP transition. And then, the chemical potential $\mu  = {\mu _{{\rm{C}}}}$ is fixed to study the shear viscosity coefficient components for the first order transition. It is found that ${\eta _{1,2,3,4}}/{\Lambda ^3}$ all increases with the temperature from low temperature to high temperature and then reaches the first order phase transition point.
	Each of ${\eta _{1,2,3,4}}/{\Lambda ^3}$ has a discontinuous at the phase transition point; an upward jump from the chiral broken phase to the chiral restoration phase is shown in Fig. 9.
	${\eta _{1,2,3,4}}/{\Lambda ^3}$ increasing with the temperature are very similar to that of first order phase transition, but ${\eta _{1,2,3,4}}/{\Lambda ^3}$ of CEP at the phase transition point continuously changes with temperature.
	\begin{figure}[H]
		\centering
		\includegraphics[width=0.42\textwidth]{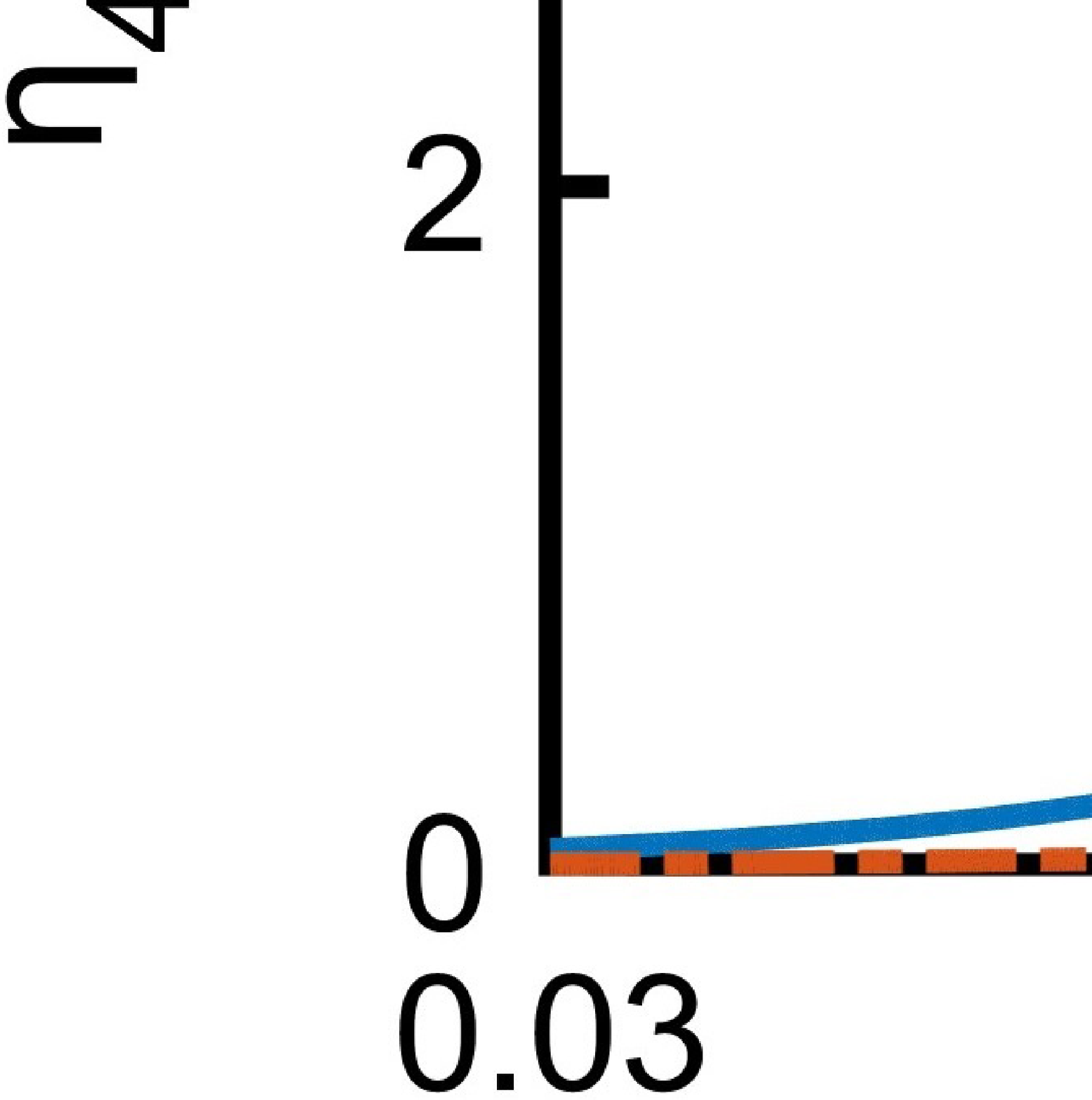}
		\caption{\label{fig10} The dependencies of renormalized shear viscosity coefficient components
			(${\eta _{1,2,3,4}}/{\Lambda ^3}$) on the temperature of first order phase transition with the magnetic fields $eB$ = 0.2 ${\rm{Ge}}{{\rm{V}}^{\rm{2}}}$ and $eB$ = 0.3 ${\rm{Ge}}{{\rm{V}}^{\rm{2}}}$, respectively.  (a) for $\eta_{1}/\Lambda^{3}$, (b) for $\eta_{2}/\Lambda^{3}$, (c) for $\eta_{3}/\Lambda^{3}$ and (d) for $\eta_{4}/\Lambda^{3}$. }
	\end{figure}
	The dependencies of shear viscosity coefficient components (${\eta _{1,2,3,4}}/{\Lambda ^3}$) on temperature of first order phase transition with magnetic fields $eB$ = 0.2 ${\rm{Ge}}{{\rm{V}}^{\rm{2}}}$ and $eB$ = 0.3 ${\rm{Ge}}{{\rm{V}}^{\rm{2}}}$ are respectively displayed in Fig. 10(a, b, c and d). The following characteristics have been revealed as:  (1) Discontinuities of ${\eta _{1,2,3,4}}/{\Lambda ^3}$ for the first order phase transition point, and there is an upward discontinuous jump from the chiral broken phase to chiral restoration phase of the high temperature; (2) ${\eta _{2}}/{\Lambda ^3}$ and ${\eta _{4}}/{\Lambda ^3}$, their jump values at the first order phase transition point increase with magnetic field; (3) Comparing ${\eta _{1,2,3,4}}/{\Lambda ^3}$, it is evident that at the same temperature,  shear viscosity coefficient components satisfy the following relationship ${\eta _4} > {\eta _3} > {\eta _2} > {\eta _1}$.
	
	\section{SUMMARY AND CONCLUSIONS}\label{sec:05 setup}
Noncentral heavy ion collisions of RHIC and LHC can produce a strong magnetic field. The phase transition under such a magnetic field during the evolution of QGP to hadron gas depends mainly on transport coefficients like shear viscosity. The dissipative coefficients in the presence of a magnetic field are also necessary and are essential ingredients for the magnetohydrodynamic evolution of the strongly interacting medium. When the magnetic field exists, the shear viscosity coefficient of the dissipative fluid system can be decomposed into five different components $\eta_0$, $\eta_1$, $\eta_2$, $\eta_3$, and $\eta_4$ as we know that $\eta_0$ is the same as the shear viscosity coefficient with zero magnetic field. We conducted a detailed study of the dependencies of $\eta_1$, $\eta_2$, $\eta_3$, and $\eta_4$ on temperature and magnetic field for the first order phase transition and critical end point transition.
	
	We have used the formalism of the NJL model with AMM in the presence of a magnetic field to describe the magnetothermodynamics of quark matter. We get a temperature and magnetic field dependent quark mass, which will be accessed to the phase space factors of $\eta_1$, $\eta_2$, $\eta_3$, and $\eta_4$. For strongly coupled RHIC or LHC collisions, the expected quark collision width $\Gamma_c$ will take a more considerable value. To describe that zone, we need a general structure of $\eta_1$, $\eta_2$, $\eta_3$, and $\eta_4$.  A detailed study of the dependencies of $\eta_1$, $\eta_2$, $\eta_3$, and $\eta_4$ on temperature and magnetic field for the first order phase transition, critical end point transition and crossover phase transition are respectively investigated.
	
	The temperature dependence of relaxation time from a simple contact diagram of $2\to 2$ scattering processes is analyzed by using the interaction Lagrangian density of the NJL model with AMM. We compared the relaxation time with and without the introduction of AMM. It is found that AMM can make the minimum point of $\tau$ more apparent and the increased amplitude more significant under high temperature conditions. Both the ratio $\eta/s$ of shear viscosity coefficient to entropy and the collision relaxation time $\tau$ show  similar trend with temperature, both of which reach minima around the critical temperature.
	
	The shear viscosity coefficient components $\eta_1$, $\eta_2$, $\eta_3$, and $\eta_4$ all increase with temperature. Discontinuities of $\eta_1$, $\eta_2$, $\eta_3$, and $\eta_4$ for the first order phase transition point, and there is an upward discontinuous jump from the chiral broken phase of the low temperature to the chiral restoration phase of the high temperature. That implies the shear viscosity coefficient of magnetized QCD medium is not a function of smooth transition for the first order phase transition.
	
	In this paper, we only discuss the viscosity coefficient by using the two-flavors NJL model with AMM. It is feasible to analyze further the influence of high-order meson fluctuations in scattering amplitude on the dissipation characteristics of QCD fluids. In the near future, we intend to approach more calculations of scattering diagrams in the 2$\rightarrow$2 process in the presence of the magnetic field.

	\section*{Acknowledgments}
	This work was supported by the National Natural Science Foundation of China (Grants No. 11875178, No. 11475068, No. 11747115).
	
	\section*{References}
	
	\bibliography{ref}
\end{document}